**Dataset and Methodology for Material Identification Using AFM Phase Approach Curves**

*Stefan R. Anton[1], Denis E. Tranca[1], Stefan G. Stanciu[1], Adrian M. Ionescu[2]\* and George A. Stanciu[1]\**

[1] Center for Microscopy-Microanalysis and Information Processing, National University for Science and Technology Politehnica of Bucharest, Bucharest, Romania

[2] Swiss Federal Institute of Technology of Lausanne, Lausanne, Switzerland

\* Corresponding authors e-mail: adrian.ionescu@epfl.ch, stanciu@physics.pub.ro

**Abstract**

Atomic force microscopy (AFM) phase approach-curves have significant potential for nanoscale material characterization, however, the availability of robust datasets and automated analysis tools has been limited. In this paper, we introduce a novel methodology for material identification using a high-dimensional dataset consisting of AFM phase approach-curves collected from five distinct materials: silicon, silicon dioxide, platinum, silver, and gold. Each measurement comprises 50 phase values obtained at progressively increasing tip-sample distances, resulting in 50×50×50 voxel images that represent phase variations at different depths. Using this dataset, we compare k-nearest neighbors (KNN), random forest (RF), and feedforward neural network (FNN) methods for material segmentation. Our results indicate that the FNN provides the highest accuracy and F1 score, outperforming more traditional approaches. Finally, we demonstrate the practical value of these segmented maps by generating simulated scattering-type scanning near-field optical microscopy (s-SNOM) images, highlighting how AFM phase approach-curves can be leveraged to produce detailed, predictive tools for nanoscale optical analysis.

**Introduction**

Atomic force microscopy (AFM) has become an indispensable tool in nanotechnology and materials science, providing unparalleled insights into surface properties and nanoscale interactions [1], [2], [3]. Together with related imaging techniques, AFM can precisely measure mechanical [4], [5], [6], electrical [7], [8], [9] , magnetic [10] and chemical [11], [12] characteristics by analysing interactions between a nano-probe tip and material surfaces. These capabilities have made it particularly valuable in materials science [13], nanotechnology [14], and biomedical research [15].

As AFM continues to evolve, there is a growing need for well-curated datasets and advanced analytical tools capable of fully exploiting AFM's potential for material identification and characterization. Recent advancements in computational infrastructure and analytical algorithms, particularly machine learning techniques, have significantly enhanced the analysis of AFM data, including classical AFM topography images and various approach curves (force, amplitude, and phase). A growing body of research [16] has explored machine learning algorithms and automated processing techniques applied to AFM datasets, with particular attention to automated material classification [17] and semantic image segmentation [18].

One critical development in AFM data analysis has been the application of machine learning to multidimensional AFM datasets for the identification of various sample types. For instance, Dokukin et al. [19] introduced the use of ensemble machine learning methods to process complex, multidimensional AFM data, specifically employing dimensionality reduction techniques and a subset of key parameters for data processing. Their work demonstrated that dimensionality reduction could improve classification efficiency and accuracy, setting a foundation for applying machine learning methods with large AFM datasets to differentiate material types. However, their study focused purely on features extracted from topography data, leaving a gap in methods based on approach-curves investigations. Another relevant study [20], reported the use of feedforward neural networks into AFM force-distance curves to diagnose cancer tissue samples automatically. The study showed that AFM data could be classified with high accuracy without extensive manual intervention by leveraging neural networks for force curve analysis. However, the cited work was limited to binary classification (cancerous vs. healthy tissue) and thus does not address the segmentation challenge of multiple materials with distinct AFM phase response profiles as in our dataset.

Moreover, recent studies [21], [22] have demonstrated that AFM phase data, which provides distinct insights into surface properties, can be highly effective in identifying material compositions. For instance, Lozano et al. [23] developed a theoretical framework for phase spectroscopy in bimodal AFM, showing that AFM phase shifts correlate with compositional variations under small conservative forces. Their work illustrated the sensitivity of AFM phase data to material-specific properties, underscoring its potential for high-fidelity material segmentation in complex samples. This theoretical basis supports phase data analysis in AFM, justifying the use of phase approach curves for semantic segmentation, as proposed in the current study.

These prior studies collectively establish the potential and growing interest in machine learning and advanced data processing in AFM applications. However, they also underscore existing limitations, including a focus on binary classification, low sample diversity, or specialized analysis modes.

This paper introduces a novel high-dimensional dataset of AFM phase approach-curves collected across five distinct materials: silicon (Si), silicon dioxide ($SiO_2$), silver (Ag), platinum (Pt), and gold (Au), each of which exhibits unique material characteristics in response to AFM probing. The generated AFM phase approach curves provide a powerful basis for material characterization and differentiation, as each material responds uniquely to interactions with the AFM probe. This diversity in response creates a rich dataset that can be leveraged to identify materials at high spatial resolution.

Recognizing the need for advanced data processing in such multidimensional datasets, this study also explores a range of classical and machine learning techniques to facilitate automated semantic segmentation of the AFM images. The resulting segmented images are more than simple visual representations, they hold further potential for simulating scattering-type scanning near-field optical microscopy (s-SNOM) images, a powerful imaging technique for visualizing nanoscale optical properties and material composition [24]. By further generating simulated s-SNOM images from the segmented AFM data, this study illustrates how AFM phase approach-curves can be used not only for material identification but also as a low-cost alternative for predicting material response under different imaging techniques, such as s-SNOM.

## Materials and Methods

### Imaging tool

The imaging process utilized the Neaspec neaSNOM 3D acquisition module, designed for capturing volumetric data. All samples were scanned using a gold-coated AFM tip (MikroMasch, HQ:NSC10/Cr-Au) to ensure consistent interaction between the tip and sample surfaces. Due to practical time constraints, the resolution was set at 50×50×50 pixels per scan, covering a physical volume of 500×500×20 nm. This high-resolution sampling approach captures detailed phase information in both lateral and vertical directions, enabling precise segmentation and accurate material differentiation at the nanometer scale.

### Samples

For the CS-20NG xyz calibration nanogrid sample, referred to here as "$SiO_2$ on Si", images were captured specifically at the boundary between silicon (Si) and silicon dioxide ($SiO_2$) regions. The topography image clearly shows a sharp interface between these two materials, allowing manual segmentation to create accurate masks that define each region. These manually generated masks serve as ground-truth references for validating algorithms designed to differentiate materials based on variations in the AFM phase signal. This sample is particularly valuable for testing segmentation methods, as it provides a well-defined boundary between two materials frequently analyzed in scanning probe microscopy.

In contrast, the "Pt on Si" sample (platinum domains deposited on silicon substrate) and "Ag on Si" sample (silver domains deposited on silicon substrate) show diffuse or gradual transitions at material boundaries, making them unsuitable for interface-based segmentation. To overcome this limitation, images were instead acquired from separate, homogeneous regions of Pt, Ag, and Si, ensuring each image set can be reliably associated with a single, known material.

Lastly, the "Au" sample consists of the gold layer coating the chip of an Au-coated AFM cantilever (MikroMasch, HQ:NSC10/Cr-Au). Images for this sample were acquired from the flat region of the cantilever surface.

### Techniques

The classification techniques evaluated in this study are k-nearest neighbors (KNN), random forest (RF), and feedforward neural networks (FNN).

K-nearest neighbors (KNN) is a distance-based classification method that delays decision-making until a new sample is introduced [25]. At that point, KNN identifies the nearest neighbors within the training dataset and assigns the class label most common among them. Although simple and intuitive, KNN can become computationally intensive with large datasets and may require careful feature selection to maintain accuracy, especially when the number of features is high.

Random Forest (RF) [26] is an ensemble method that combines predictions from multiple decision trees. Each tree is trained on a different bootstrap sample drawn from the original data, and random subsets of features are considered at each decision split. By aggregating the outcomes of many trees, Random Forest reduces variance and generally achieves robust performance, also allowing estimation of feature importance.

Feedforward Neural Networks (FNN) [27] consist of interconnected layers of artificial neurons that learn complex data patterns through a process called backpropagation. Each neuron applies a weighted sum to its inputs and then transforms the result using an activation function. This architecture enables the network to model complex, non-linear relationships when provided with sufficient training data and appropriate regularization.

These algorithms can be applied directly to the original 50×50×50 voxel dataset or combined with dimensionality reduction methods. By classifying approach curves based on their phase profiles, each voxel is labeled with its predicted material. This approach enables the creation of segmented maps highlighting distinct regions within the samples analyzed.

**Results and discussion**

*Dataset Description*

The dataset was obtained from three-dimensional AFM scans, recorded as 50×50×50 voxel images spanning a physical volume of 500×500×20 nm. Each voxel corresponds to an individual phase measurement taken at incremental points along the Z-axis, resulting in a detailed phase response profile for every XY coordinate.

The dataset encompasses a diverse collection of images from multiple materials. Table 1 summarizes the number of images acquired for each sample, totaling 65 images. For the "$SiO_2$ on Si" sample, a total of 21 regions were imaged, each containing both Si and $SiO_2$ areas. For the "Pt on Si" sample, 15 regions were analyzed, including 13 images of Pt and 2 images of Si. The "Ag on Si" sample includes a total of 24 regions, comprising 14 images of Ag and 10 images of Si. Finally, for the "Au" sample, 5 separate regions were scanned.

Table 1: Number of image regions in the dataset for each sample, representing five different materials.

| Sample | Material | No. of imaged regions |
|---|---|---|
| $SiO_2$ on Si | $SiO_2$/Si | 21 |
| Pt on Si | Pt/Si | 15 [13 Pt, 2 Si] |
| Ag on Si | Ag/Si | 24 [14 Ag, 10 Si] |
| Au | Au | 5 |

The phase data acquisition at each XY pixel produces a vector containing 50 phase values. Each vector captures the phase variation as the AFM tip retracts from direct contact with the sample surface up to a maximum separation height of 20 nm. Because the phase signal can show discontinuities due to phase-wrapping, all phase vectors undergo an unwrapping procedure to ensure smooth and continuous measurements across the entire range. After unwrapping, the final value of each phase vector, corresponding to the maximum tip-sample separation, is subtracted from every element of the vector. This step standardizes the phase signals by setting a baseline of zero when the tip is furthest from the sample. This process effectively removes any intrinsic phase contribution from the AFM tip itself, isolating the material-specific response.

Using this procedure, we compiled a comprehensive dataset of phase curves (summarized in Table 2), enabling detailed analysis for each material. The total number of phase curves collected for each material type is presented in the table below.

Table 2: Number of phase curves in the dataset associated with each of the five materials.

| Material | No. of phase curves |
|---|---|
| Si | 58133 |
| SiO2 | 24367 |
| Pt | 32500 |
| Ag | 35000 |
| Au | 12500 |

Each phase curve entry in the dataset is accompanied by additional metadata, including labels for "Area" (also referred to as "Image Index"), "Sample," and "Material." The "Area" column provides a unique index (ranging from 1 to 65) for each original image, enabling quick retrieval and reconstruction of the corresponding 50×50×50 voxel 3D image. The "Sample" column identifies the sample type (e.g., "SiO$_2$ on Si," "Pt on Si," etc.), while the "Material" column specifies the material associated with each phase curve (Si, SiO$_2$, Pt, Ag, or Au). Columns numbered from 1 to 50 contain the measured phase values recorded as the AFM tip retracts from the sample surface.

An example of a 3D phase image reconstructed from this dataset is shown in the following figure. The contrast between Si and SiO$_2$ regions is clearly distinguishable at various tip-sample separations. The differences in contrast observed at different heights highlight the depth-dependent variations in phase response, providing valuable information that can significantly improve material classification.

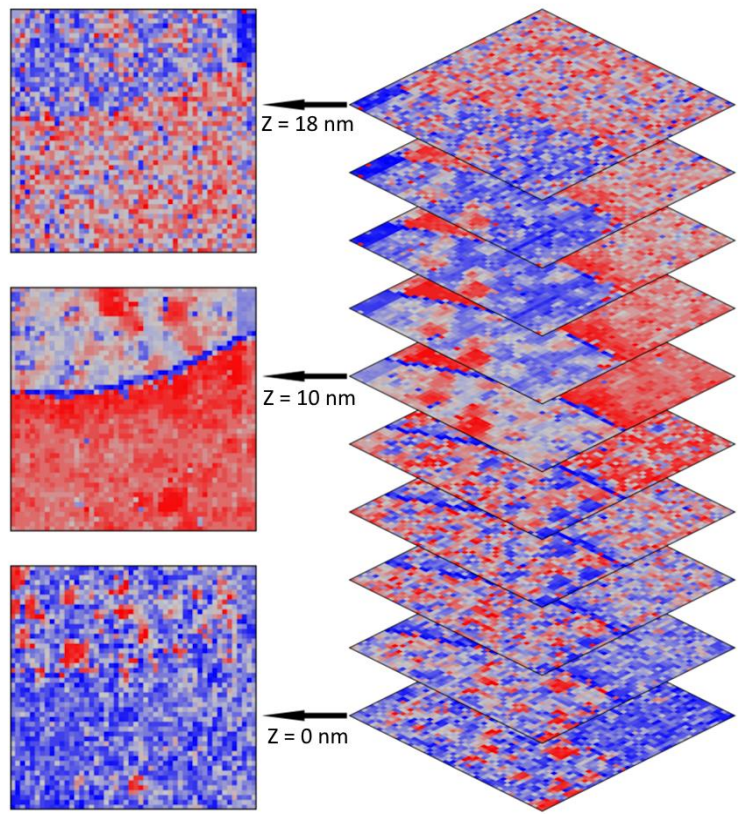

Figure 1: AFM phase image stack generated from a single sample from the dataset.

While phase contrast is generally noticeable across different materials, certain material pairs exhibit overlapping phase values. For instance, the following figure presents the mean phase per material (solid lines) along with the standard deviation (shaded regions). It demonstrates significant overlap between the phase signals of Si and $SiO_2$, regardless of the tip-sample separation distance.

For Au and Pt, the phase curves are similar when the tip is in contact with the surface. However, as the tip retracts, their phase signals begin to diverge, with Au displaying a noticeably lower phase shift compared to Pt at separations of approximately 5–10 nm. This divergence reflects distinct material properties that can be leveraged in material identification strategies based on the full phase-distance curves.

The overlaps observed in the phase signals emphasize the importance of analyzing complete phase-distance curves rather than relying solely on phase measurements obtained at a single tip-sample distance. This comprehensive approach provides critical information necessary for accurate material segmentation.

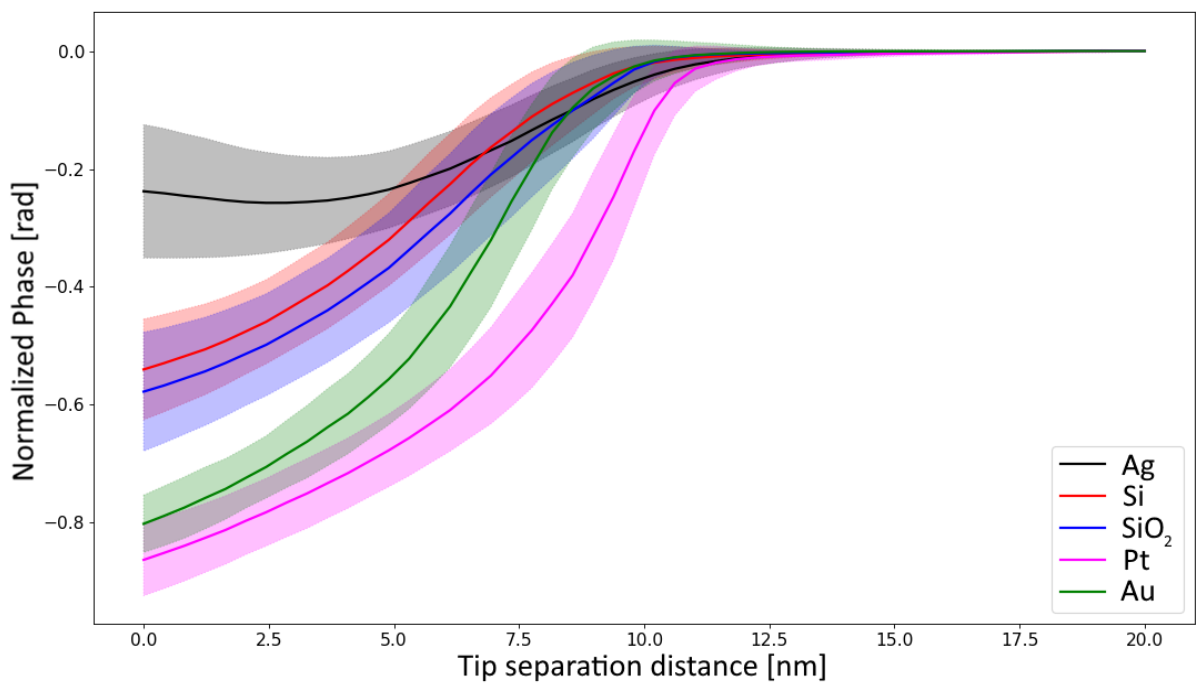

Figure 2: Mean (solid lines) and standard deviation (shaded regions) of the dataset entries grouped by materials: Ag (black), Si (red), $SiO_2$ (blue), Pt (magenta), and Au (green).

***Dimensionality reduction***

Although the dataset is information-rich due to its high-dimensional structure, it is challenging to analyze because of its large size and relatively low signal-to-noise ratio.

The stochastic nature of the measurements [28], combined with the need to process many phase curves to obtain statistically reliable results, complicates manual analysis. To address this challenge, we utilize dimensionality reduction techniques to identify a small set of key parameters from each phase curve, enabling more efficient material segmentation.

We applied several dimensionality reduction methods, including Principal Component Analysis (PCA) [29], Independent Component Analysis (ICA) [30] , and Uniform Manifold Approximation and Projection (UMAP) [31], to visualize the dataset clearly in a simplified, low-dimensional space. Additionally, we conducted a parameter sweep for UMAP, adjusting factors such as distance metrics, the number of neighbors, and minimum distance, to optimize the results.

Figure 3 shows orthographic projections resulting from PCA and ICA methods. Points from one material class appearing within clusters of another class indicate regions of overlapping phase curves, reflecting the dataset's inherent low signal-to-noise ratio and material similarities.

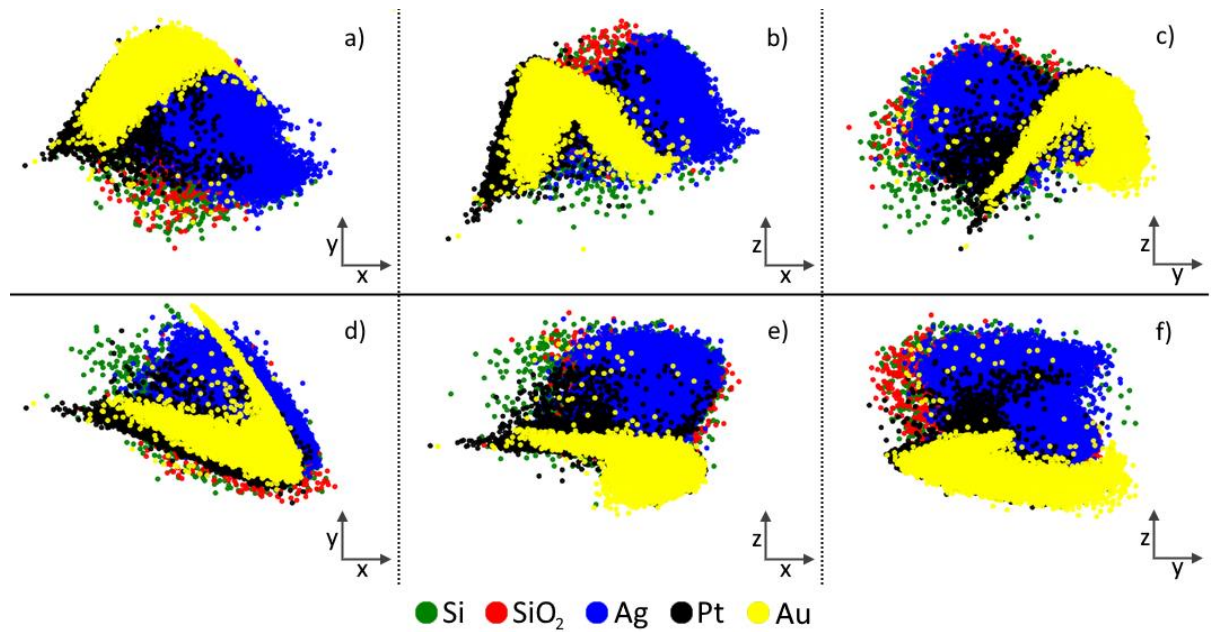

Figure 3: Two-dimensional projections of the three-dimensional feature space obtained by applying PCA and ICA to the original dataset. Panels (a), (b), and (c) represent PCA projections, while panels (d), (e), and (f) show ICA projections. Specifically, (a) and (d) illustrate projections onto the XY-plane, (b) and (e) onto the XZ-plane, and (c) and (f) onto the YZ-plane.

*Evaluation dataset preparation*

To evaluate and compare the methods presented in this study, we constructed a dedicated test dataset consisting of one 3D phase image for each material type across four distinct samples: Si/SiO$_2$, Pt, Ag, and Au. For Pt, Ag, and Au, the segmentation masks corresponded to uniform material regions, whereas the SiO$_2$ on Si sample contained two clearly defined regions around the Si/SiO$_2$ interface.

The training and validation datasets were created by dividing the remaining data (excluding the test set) into two subsets, with proportions of 75% for training and 25% for validation. This ensured the test dataset contained only phase profiles from the four specified images, making it completely independent of the training and validation sets. Such clear separation is essential to mimic real-world scenarios, where models must accurately segment new, unseen data. Importantly, the test dataset was not used during hyperparameter tuning or intermediate evaluations to avoid biasing the results.

**K-Nearest Neighbours (KNN)**

For the KNN model, the main hyperparameter of interest was the number of neighbors. To determine the optimal value, we conducted an exhaustive search ranging from 2 to 5000 neighbors. The optimal neighbor count was selected based on classification accuracy measured on the validation dataset. Additionally, we improved classification performance by implementing a majority-vote ensemble. This ensemble approach combined predictions from the top five configurations ranked by their F1 scores (harmonic mean of precision and recall), effectively combining the strengths of multiple models to enhance segmentation accuracy.

Table 3 summarizes the performance of the various configurations evaluated. The table lists the dimensionality reduction methods alongside their corresponding accuracy and F1 scores on the validation dataset. For UMAP projections, only the top five configurations based on F1 scores are presented.

We also evaluated dimensionality reduction techniques such as PCA and ICA separately. These methods achieved moderate success, with F1 scores of 0.6867 (PCA) and 0.6882 (ICA). Although dimensionality reduction methods reduced computational complexity, their performance was slightly below the model using the raw features. The highest performance was obtained using the majority-vote ensemble of the top five dimensionality-reduction-based configurations, achieving an accuracy of 79.66% and an F1 score of 0.7639.

Table 3: Accuracy and F1-score results for segmentation using the KNN model under various configurations. The best performing configuration is bolded.

| Method | Accuracy [%] | F1 Score |
| --- | --- | --- |
| Manhattan (neighbours = 10, min_dist = 0.05) | 78.85 | 0.7529 |
| Manhattan (neighbours = 10, min_dist = 0.25) | 78.86 | 0.7529 |
| Canberra (neighbours = 25, min_dist = 1.0) | 78.65 | 0.7517 |
| Manhattan (neighbours = 25, min_dist = 0.25) | 78.70 | 0.7516 |
| Canberra (neighbours = 25, min_dist = 0.25) | 78.54 | 0.7514 |
| PCA | 73.36 | 0.6867 |
| ICA | 73.55 | 0.6882 |
| **Majority vote** | **79.66** | **0.7639** |
| All features | 76.62 | 0.7333 |

The segmentation results obtained using the majority-vote ensemble on the independent test set are shown in Figure 4. Although this ensemble achieved the highest overall performance metrics, visual analysis revealed some notable limitations. The segmented images exhibited considerable noise, especially within homogeneous regions. In the case of the Si/$SiO_2$ sample, the segmentation of the $SiO_2$ region was particularly problematic, with extensive misclassification across the image. These inaccuracies highlight the inherent challenges of using KNN for accurate segmentation tasks.

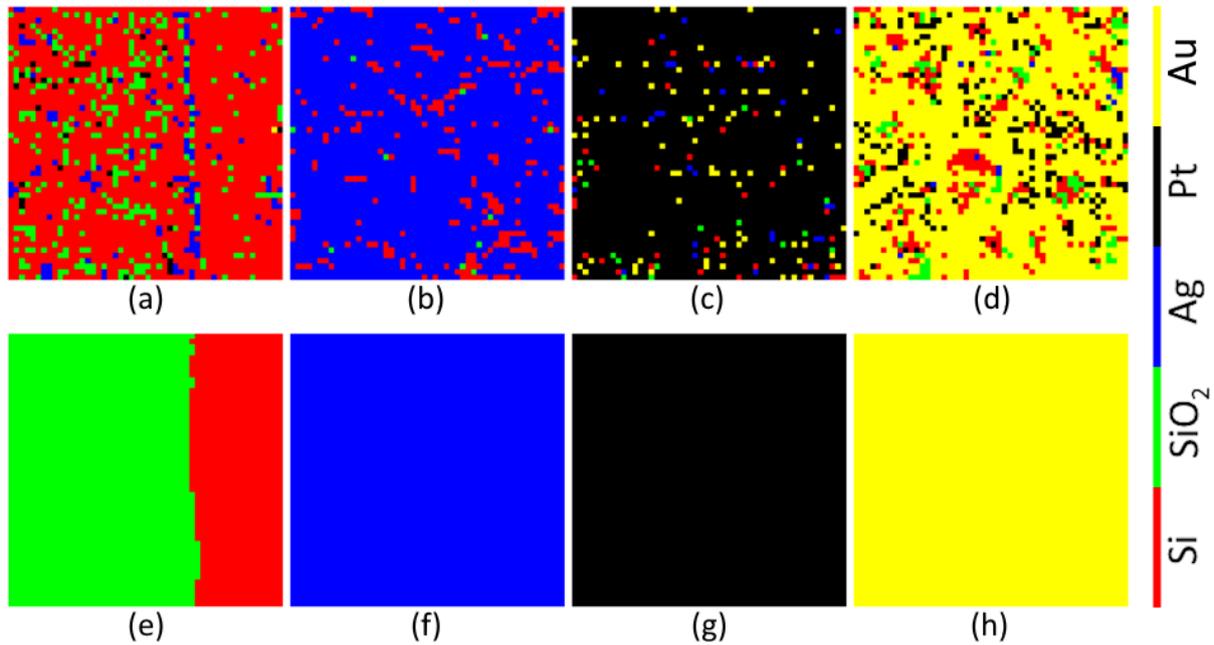

Figure 4: Segmentation masks for the test images obtained using the KNN model (a–d) compared with manual segmentation (e–h). The materials and their corresponding color codes are: (a, e) Silicon (red) and Silicon Dioxide (green); (b, f) Silver (blue); (c, g) Platinum (black); and (d, h) Gold (yellow).

**Random Forest (RF)**

The Random Forest (RF) model was evaluated using the same data-splitting strategy as the KNN method. A crucial hyperparameter for the RF model is the number of estimators, which determines how many decision trees are included in the ensemble. Typically, increasing the number of estimators improves classification accuracy, though the benefits diminish as more trees are added. To balance accuracy with computational efficiency, we chose the optimal number of estimators based on the following criterion: the improvement in accuracy between consecutive values of estimators must remain below 0.01% for at least 10 consecutive values. This approach ensured the selection of a stable and efficient model.

As with the KNN approach, we implemented a majority-vote ensemble strategy to enhance segmentation accuracy. This ensemble combined predictions from the top five configurations (ranked by F1 score), leveraging their complementary strengths for improved accuracy and robustness.

The performance results of the Random Forest models are summarized in Table 4, showing accuracy and F1 scores computed on the validation dataset. Results include classical dimensionality reduction methods (PCA and ICA), the raw-feature-based predictions, and the majority-vote ensemble.

The best individual dimensionality reduction model (UMAP) achieved an accuracy of 82.61% and an F1 score of 0.8059. Classical methods such as PCA and ICA yielded slightly lower performance, with accuracy values of 79.80% and 79.64%, respectively. However, the majority-vote ensemble combining the top five UMAP configurations improved performance, achieving an accuracy of 84.56% and an F1 score of 0.8149, demonstrating the advantage of combining multiple classifiers. Ultimately, the highest overall performance was obtained by using all available features directly, reaching an accuracy of 87.74% and an F1 score of 0.8635.

Table 4: Accuracy and F1-score results for segmentation using the Random Forest model across various configurations. The best performing configuration is bolded.

| Method | Accuracy | F1 Score |
| --- | --- | --- |
| Manhattan (neighbours = 10, min_dist = 0.05) | 82.61 | 0.80592 |
| Manhattan (neighbours = 10, min_dist = 0.25) | 82.44 | 0.804335 |
| Bray-Curtis (neighbours = 10, min_dist = 0.05) | 82.43 | 0.802404 |
| Manhattan (neighbours = 25, min_dist = 0.05) | 82.25 | 0.801466 |
| Bray-Curtis (neighbours = 10, min_dist = 0.25) | 82.36 | 0.800988 |
| PCA | 79.80 | 0.7287 |
| ICA | 79.64 | 0.7267 |
| Majority vote | 83.56 | 0.8149 |
| **All features** | **87.74** | **0.8635** |

The segmentation masks generated by the RF model using all available features (Figure 5) showed notable improvements compared to those obtained with the KNN method. Specifically, uniform regions in the segmentation masks (such as Pt, Ag, and Au) displayed significantly reduced noise, reflecting better discrimination between materials. Additionally, segmentation accuracy for the Si/SiO$_2$ sample was substantially improved, resulting in a mostly correct identification of the SiO$_2$ region.

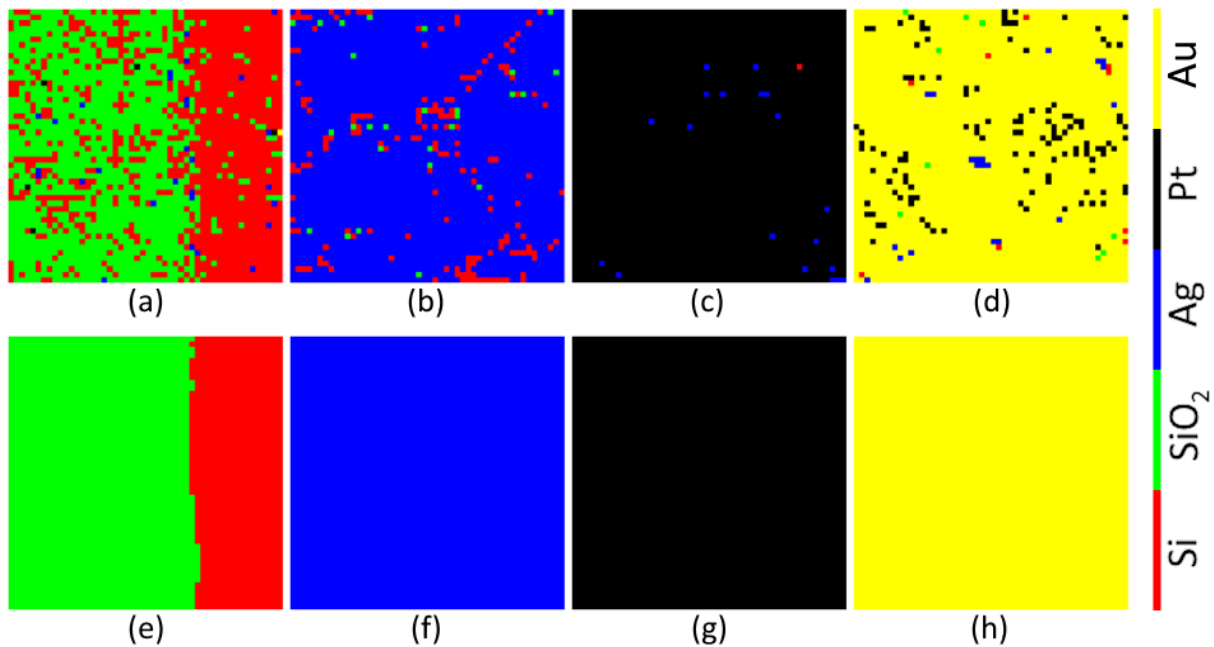

Figure 5: Segmentation masks for the test images obtained using the RF model (a–d) compared with manual segmentation (e–h). The materials and their corresponding color codes are: (a, e)

Silicon (red) and Silicon Dioxide (green); (b, f) Silver (blue); (c, g) Platinum (black); and (d, h) Gold (yellow).

The improved performance of the Random Forest models can be attributed to their ensemble-based design, which inherently reduces the influence of noise and data variability. By averaging predictions from multiple decision trees, Random Forest achieves more robust and stable segmentation results. However, despite these significant improvements, certain challenges remain. Specifically, accurately segmenting fine-grained features, such as the precise boundary between Si and $SiO_2$, could still be enhanced through further optimization.

**Feedforward neural networks (FNN)**

The FNN architecture was carefully designed to balance complexity and training stability. The goal was to create a network deep enough to effectively capture complex patterns in the dataset, while avoiding training issues such as vanishing gradients that commonly affect deeper models. After extensive experimentation, we selected an architecture consisting of five layers with dimensions of [128, 64, 32, 16, 5], providing an effective balance between representational power and computational efficiency.

To prevent the issue of vanishing gradients, we used the Rectified Linear Unit (ReLU) activation function. Batch normalization was also incorporated, as it stabilizes training by normalizing activations within each layer, thus reducing internal covariate shifts. The model was trained for 50 epochs using the complete set of features, without applying any dimensionality reduction. This ensured that the neural network had access to all available information, maximizing its capability to learn complex relationships in the data.

The FNN achieved the highest accuracy and F1 score among all tested models. The performance of the best-performing configurations is summarized in Table 5. The notable improvement in both accuracy and F1 score demonstrates the model's ability to effectively handle dataset complexity, with the high F1 score indicating balanced and robust performance across all materials.

Table 5: Accuracy and F1-score results for segmentation using the FNN model across various configurations, including variations in the number of layers, neurons per layer, and activation functions. The best performing configuration is bolded.

| Method | Accuracy | F1 Score |
| --- | --- | --- |
| [128, 64, 32, 16, 5] – ReLu | **0.9357** | **0.9064** |
| [128, 64, 32, 16, 5] - Tanh | 0.9278 | 0.9019 |
| [128,64,32,5] - ReLu | 0.9112 | 0.8746 |
| [128,64,32,5] - Tanh | 0.8892 | 0.8590 |
| [128,53,5] - Tanh | 0.8758 | 0.8561 |
| [128,53,5] - ReLU | 0.8559 | 0.8109 |

The segmentation masks produced by the FNN (Figure 6) show clear improvements in quality compared to those generated by the KNN and RF models. For uniform material regions, the noise levels in FNN masks are similar to those obtained with the RF model. However, the mask for the $SiO_2$ on Si sample shows significantly improved accuracy, especially along the material boundary. This enhanced performance results from utilizing the complete feature set without dimensionality reduction, thus preventing information loss. Additionally, the selected neural network architecture, combined with the ReLU activation function and batch normalization, provided robust training conditions that minimized common issues such as overfitting.

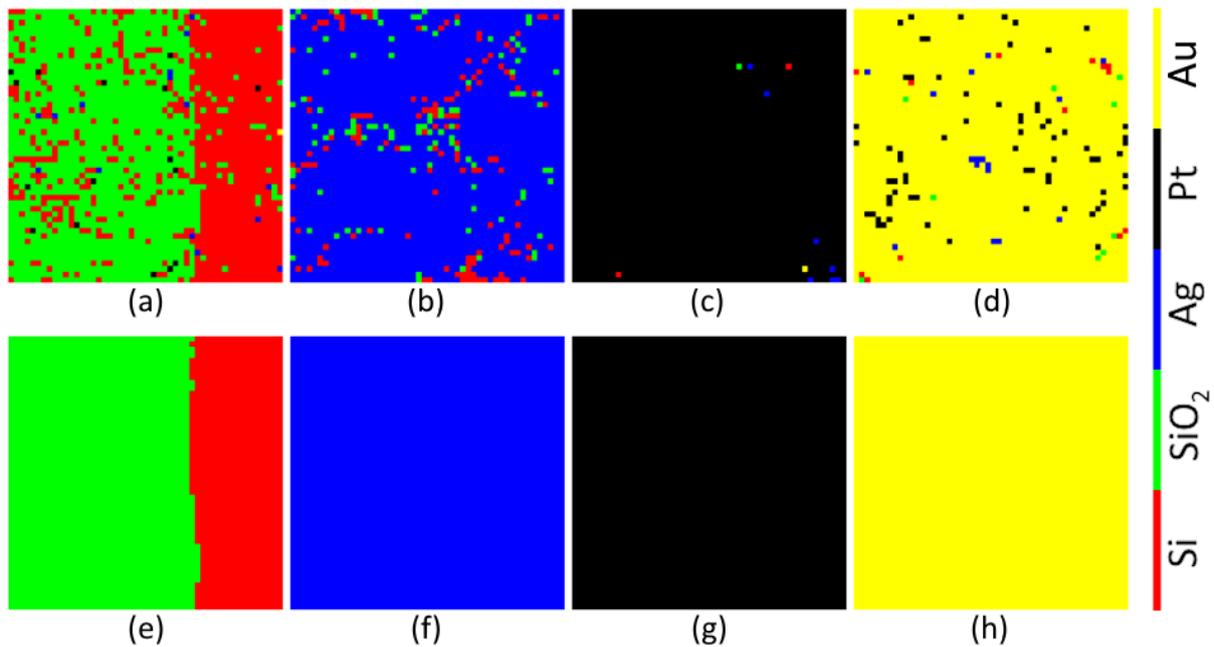

Figure 6: Segmentation masks for the test images obtained using the FNN model (a–d) compared with manual segmentation (e–h). The materials and their corresponding color codes are: (a, e) Silicon (red) and Silicon Dioxide (green); (b, f) Silver (blue); (c, g) Platinum (black); and (d, h) Gold (yellow).

Compared to KNN and RF, the FNN model achieved the highest accuracy and F1 score and produced the most visually consistent segmentation masks. The simplicity and effectiveness of the FNN architecture make it particularly well-suited for analyzing high-dimensional data. However, FNN models required greater computational resources, especially during the training phase, compared to simpler methods such as KNN.

**Simulated s-SNOM**

In this section, we investigate the feasibility of generating simulated s-SNOM images using a three-dimensional phase image from our dataset. To achieve this, we use the best-performing FNN model identified earlier to assign material labels to each pixel, converting the 3D phase data into a segmented 2D map that visualizes the spatial distribution of materials within the sample.

From this segmented image, we simulate the s-SNOM amplitude at each pixel using known optical parameters for the identified materials. The simulation is performed using the snompy package [32], with material parameters selected at a wavelength of 1550 nm to match experimental conditions.

In addition to generating a simulated s-SNOM amplitude image directly from the raw segmentation mask, we apply median filtering to the mask before simulation. This filtering step helps reduce localized classification errors and suppress noise artifacts resulting from segmentation inaccuracies. By smoothing small, isolated misclassified regions, median filtering provides a cleaner and more accurate representation of material distributions, thus improving the simulated s-SNOM response.

Figure 7 shows a side-by-side comparison of the experimental s-SNOM image and the two simulated images: one generated from the raw segmentation mask, and the other from the median-filtered mask. This visual comparison provides an initial qualitative assessment of how closely the simulations match experimental observations.

To support this visual evaluation, Table 6 presents a quantitative comparison between experimental data and simulations. The results indicate that the simulated s-SNOM image based on the median-filtered mask more accurately aligns with experimental measurements compared to the unfiltered mask. These findings emphasize the effectiveness of incorporating simple post-processing steps like median filtering, reinforcing the reliability and accuracy of our segmentation and simulation pipeline.

Table 6: Quantitative comparison between experimental and simulated s-SNOM images.

| Material | Experimental average amplitude | Simulated average amplitude | Simulated average amplitude median filter |
|---|---|---|---|
| Si | 0.8373 ± 0.1022 | 0.8198 ± 0.0509 | 0.8202 ± 0.0300 |
| $SiO_2$ | 0.7367 ± 0.1128 | 0.7932 ± 0.1383 | 0.7191 ± 0.0982 |

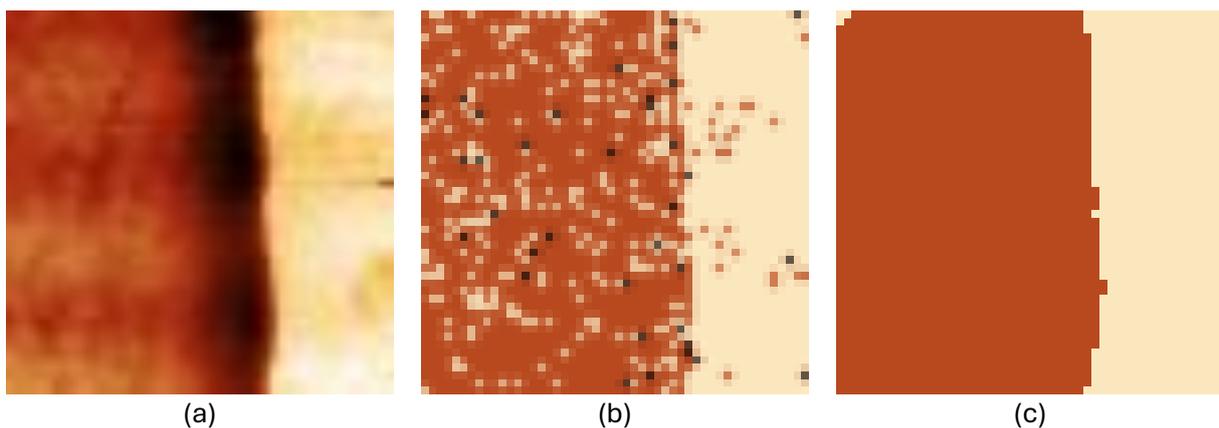

(a) (b) (c)

Figure 7: (a) Experimental s-SNOM amplitude image showing the boundary region between Si and $SiO_2$, (b) Simulated s-SNOM amplitude image generated from the segmentation mask obtained using the FNN model and snompy, (c) Simulated s-SNOM amplitude image generated from the median-filtered segmentation mask obtained using the FNN model and snompy.

**Conclusion**

This study presented a comprehensive dataset of three-dimensional AFM phase approach curves, showcasing the potential of machine learning to enhance nanoscale material characterization. By leveraging high-dimensional phase data from five representative materials (Si, $SiO_2$, Pt, Ag, and Au) and comparing multiple classification methods (k-nearest neighbors, random forest, and feedforward neural networks), we demonstrated that neural networks excel at learning non-linear relationships, resulting in superior accuracy and reliability for material segmentation.

A key outcome was the successful generation of simulated s-SNOM images from the FNN-based segmentation masks. By correlating material labels with optical constants at a given wavelength, we simulated near-field optical responses that closely matched experimentally obtained s-SNOM data. This not only validates our AFM-driven segmentation approach but also highlights the promise of bridging AFM data and advanced optical simulations to predict near-field behavior without extensive or costly experiments.

Beyond immediate findings, our results underscore a growing synergy between AFM and machine learning. Traditional AFM analysis often relies on time-consuming manual interpretation, whereas automated data-driven methods can streamline workflows, improve reproducibility, and unlock new opportunities for real-time or high-throughput research. Looking ahead, expanding the dataset to include more materials, exploring advanced neural architectures, and integrating these methods with automated experimental setups could further enhance AFM's role as a quantitative platform for nanoscale discoveries in fields like nano-photonics, semiconductor manufacturing, and materials engineering.

**Funding**

This work was supported by the HORIZON-HLTH-2023-TOOL-05, Project 101137466 (RealCare). The use of the Neaspec NeaSNOM Microscope was possible due to the European Regional Development Fund through Competitiveness Operational Program 2014–2020, Priority axis 1, Project No. P_36_611, MySMIS code 107066-INOVABIOMED.